\documentclass[12pt]{article}

\usepackage{epsfig}

\usepackage{fullpage}

\begin{document}

\centerline{\bf\large Non-regular (non-St\"ackel) R-separation and general modulated soliton}

\bigskip

\centerline{\bf\large of wave equation}

\bigskip

\bigskip

\centerline{\bf R. Prus\footnote{Warsaw University, Faculty of Physics, Department of Mathematical Physics, ul. Ho\.za 74, 00-682 Warsaw, Poland, email: robert.prus@fuw.edu.pl .} and A. Sym\footnote{Warsaw University, Faculty of Physics, Department of Mathematical Physics, ul. Ho\.za 74, 00-682 Warsaw, Poland, email: antoni.sym@fuw.edu.pl .}}

\bigskip

\bigskip

{\bf ABSTRACT}

\bigskip

\noindent We present a class of orthogonal non-regular in a sense of Kalnins and Miller (hence non-St\"ackel) coordinates which are R-separable in 3-dim. Helmholtz equation. One family of parametric surfaces consists of parallel Dupin cyclides, the other two consist of circular cones. This coordinate system is used to simplify derivation of Friedlander's formulae for a general "simple progressive solution of the wave equation" (modulated soliton of wave equation) in $E^3$ and to correct some errors of his paper. The extended version of the paper will be published soon.

\bigskip

\bigskip

Presumably H. Schmidt \cite{Schmidt} was the first to pose the problem of finding a general solution of the wave equation:

\begin{equation}
\Delta_3\ {\cal F} = {\cal F}_{,tt}\label{eqn:001}
\end{equation}

\bigskip

of the form:

\begin{equation}
{\cal F} (x, y, z, t) = F(x, y, z)\ G[w(x, y, z) - t]\label{eqn:002},
\end{equation}

\bigskip

where $G$ denotes an arbitrary $C^2(R)$-function. In particular (\ref{eqn:002}) is the standard plane or spherical wave. Obviously, solution (\ref{eqn:002}) can reasonably be called "general modulated soliton of wave equation". Schmidt derived the system of three partial differential equations for unknowns $F$ and $w$ which is equivalent to the problem above and which contains a nonlinear (eikonal) equation for $w$. Schmidt, however, was unable to solve this system completely

\bigskip

The problem was undertaken independently and solved completely by F. G. Friedlander in 1946 \cite{Friedlander}. His 10-page derivation contains rather tedious calculations and, unfortunately, some mistaken formulae.

\bigskip

The aim of this paper is two-fold. The mathematical aim is to present a rather non-standard orthogonal coordinate system in $E^3$ which in historical terms belongs to the three-cyclidic systems described for the first time by young Gaston Darboux \cite{Darboux1866}. See also \cite{Darboux1873}, \cite{Darboux1917} and \cite{Darboux1910}. In modern terms - within the Levi - Civita - Kalnins - Miller theory of R-separation \cite{LeviCivita}, \cite{KalninsMiller}, \cite{Miller}. the system under discussion is an example of the so called non-regular R-separable system for 3-dim. Helmholtz equation:

\begin{equation}
%\Delta_3\ f + k^2\ f = 0\label{eqn:003}.
\left( \Delta_3 + k^2 \right) f = 0\label{eqn:003}.
\end{equation}

\bigskip

The physical aim is to apply this system to greatly simplify derivation of Friedlander's formulae and to correct some of them.

\bigskip

The construction of our orthogonal coordinate systems is strictly related to the so called "J. C. Maxwell problem" of 1867 \cite{Maxwell}. This is a problem of finding two space curves such that the congruence of lines meeting these curves admits orthogonal trajectories. Maxwell proved that the two curves must be conics in perpendicular planes, related exactly in the same manner as focal conics of a standard confocal system of quadrics \cite{Gray}. Similtaneously, the orthogonal and parallel surfaces (trajectories) are the so called Dupin cyclides (surfaces of 4-th or 3-rd degree) which are together with e.g. circular cones special cases of the general Darboux - Moutard cyclides. For modern discussion of Darboux - Moutard cyclides see \cite{Takeuchi}. Notice that curvature coordinates $(u, v)$ upon any Dupin surface supplemented by distance $w$ between two parallel Dupin cyclides defines an orthogonal system in $E^3$: $u = const.$ and $v = const.$ are circular cones while $w = const.$ are Dupin cyclides. This construction defines orthogonal systems of our paper.

\bigskip

In general position a pair of focal conics is an ellipse and hyperbola. The corresponding orthogonal system $(u, v, w)$ is denoted by (e-h) and defined explicitly by:

%\begin{eqnarray}
%x = & \frac{( c w + b^2 \cos u ) \cosh v - a w \cos u}{a \cosh v - c \cos u}\label{eqn:004a},\\
%y = & \frac{a b \cosh v \sin u - b w \sin u}{a \cosh v - c \cos u}\label{eqn:004b},\\
%z = & \frac{b w \sinh v - b c \cos u \sinh v}{a \cosh v - c \cos u}\label{eqn:004c},
%\end{eqnarray}

\begin{eqnarray}
x = & \frac{b^2 \cos u \cosh v + ( c \cosh v - a \cos u) w}{a \cosh v - c \cos u}\label{eqn:004a},\\
y = & \frac{b \sin u ( a \cosh v - w )}{a \cosh v - c \cos u}\label{eqn:004b},\\
z = & \frac{b \sinh v ( w - c \cos u )}{a \cosh v - c \cos u}\label{eqn:004c},
\end{eqnarray}

\bigskip

where $a$ and $b$ are real parameters ($0 < b < a$), $c = \sqrt{a^2 - b^2}$, and $c \cos u < w < a \cosh v$.

\bigskip

%Parametric surfaces of such cyclidic coordinates are shown in figure \ref{fig:001}.
Parametric surfaces of such cyclidic coordinates are shown in Figure 1.

\bigskip

%\begin{figure}
%\begin{center}
%\epsfbox{paper.eps}
%\end{center}
%\caption{\label{fig:001}Parametric surfaces of (e-h) cyclidic orthogonal system.}
%\end{figure}

\vbox{
\vspace{0.0in}
\vbox{\vspace{0.0in} \hspace{0.0in} \hbox{\hspace{1.0in} \psfig{figure=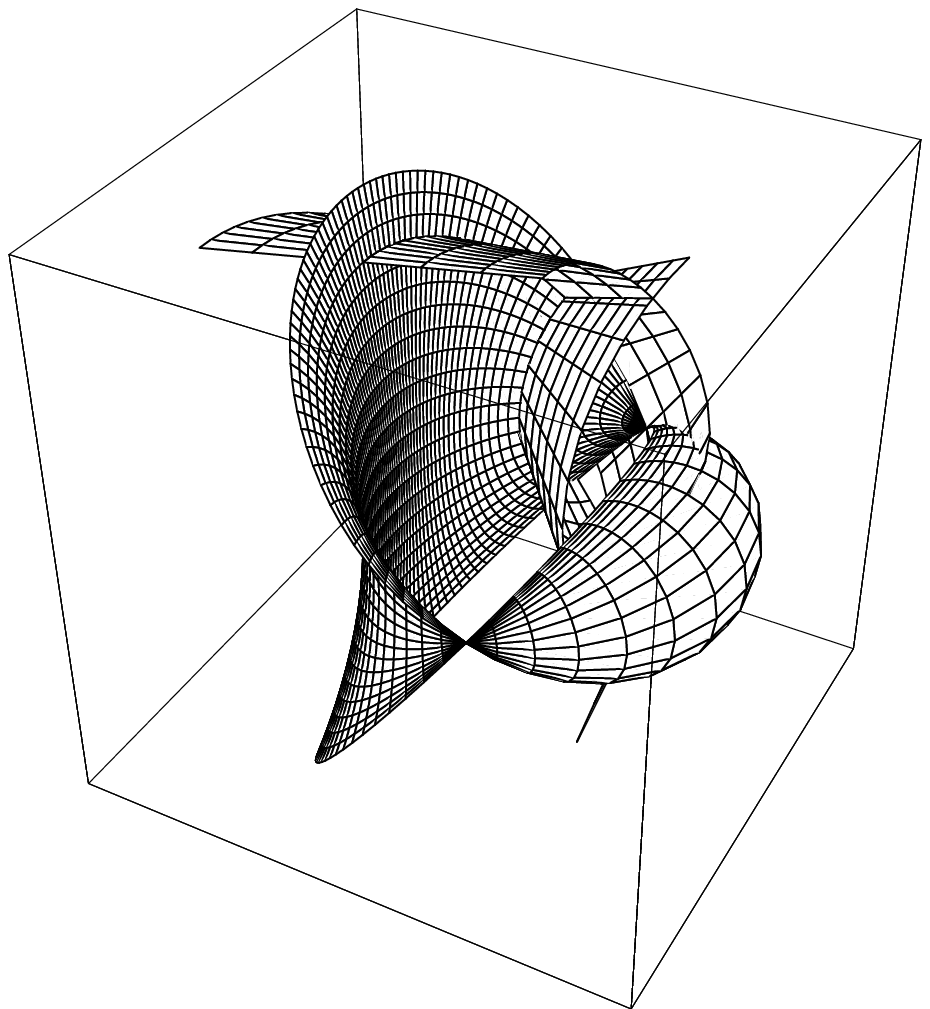,height=4in}}}
}

\bigskip

Figure 1. Parametric surfaces of (e-h) cyclidic orthogonal system.

\bigskip

When focus of hyperbola tends to infinity both focal conics degenerate into parabolas. The corresponding orthogonal system is denoted by (p-p). Finally, confluence of two ellipse foci results in a pair of a circle and straight line cutting circle orthogonally at its center.

\bigskip

{\sc Theorem 1}

\medskip

Cyclidic coordinates (e-h) are R-seperable in Helmoltz equation (\ref{eqn:003}). This means that equation (\ref{eqn:003}) rewritten in (e-h) coordinates admits solutions of the form:

\begin{equation}
f (u, v, w) = \left( w - c \cos u \right)^{-1/2} \left( a \cosh v - w \right)^{-1/2}\ U(u)\ V(v)\ W(w)\label{eqn:005},
\end{equation}

\bigskip

\noindent where $U" + \frac{1}{4}\ U = 0$, $V" - \frac{1}{4}\ V = 0$ and $W" + k^2\ W = 0$.

\bigskip

Here a universal factor in front of ordinary separation product is generally denoted by $R$. The proof of this theorem will be given in an extended version of this paper.

\bigskip

Presumably the most general setting to treat various variable separation problems is due to Kalnins - Miller \cite{KalninsMiller}, \cite{Miller} as a far reaching extension of T. Levi-Civita idea to handle the additive separation case \cite{LeviCivita}. Among many possibilities of variable separation Kalnins and Miller isolate the case of the {\it regular separation} which can be applied both to additive (mainly) and multiplicative cases (R-separation is included). E.g. the idea of regular separation (in additive case) is analytically encoded in requirement that equation (1.6) of \cite{Miller} has to be identity in derivatives of unknown. In particular, Kalnins-Miller show that orthogonal regular R-separation for Helmholtz equation (\ref{eqn:003}) necessarily leads to the metric in the so called St\"ackel form. On the other hand, according to Weinacht \cite{Weinacht} - Eisenhart \cite{Eisenhart} theorem all St\"ackel metrics in $E^3$ are those of confocal quadrics or of their appropriate degenerations. Hence any system discussed in this paper is by no means in St\"ackel form and this means is {\it non-regular}. Indeed, regular separation admits separable solutions with maximal number of parameters (confront our {\sc Theorem 1}). Thus our cyclidic systems can be hardly applied to solve typical problems of mathematical physics like boundary problems. Suprisingly, one can effectively apply our systems to re-derive Friedlander's results.

\bigskip

{\sc Theorem 2}

\medskip

The subclass of modulated solitons of wave equation (\ref{eqn:001}) in (e-h) coordinates is given by:

\begin{equation}
{\cal F} (u, v, w, t) = \left( w - c \cos u \right)^{-1/2} \left( a \cosh v - w \right)^{-1/2}\ U(u)\ V(v)\ G(w - t)\label{eqn:006}
\end{equation}

\bigskip

{\sc Proof}

\bigskip

Certainly, due to {\sc Theorem 1}:

\begin{equation}
{\cal F}_{\xi} = \left( w - c \cos u \right)^{-1/2} \left( a \cosh v - w \right)^{-1/2}\ U(u)\ V(v)\ \exp \left( -i \xi (w - t) \right)\label{eqn:007}
\end{equation}

\bigskip

where $\xi \in R$, is a solution of equation (\ref{eqn:001}).

\bigskip

Let us select any $\varphi(\xi)$ which together with $\xi \varphi(\xi)$ and $\xi^2 \varphi(\xi)$ are absolutely integrable. On performing Fourier transformation of both sides of (\ref{eqn:007}) we arrive at (\ref{eqn:006}) where:

\begin{equation}
G(w - t) = \int_{-\infty}^{\infty} \varphi(\xi)\ \exp \left( -i \xi (w - t) \right)\ d\xi\label{eqn:008}
\end{equation}

\bigskip

This is, modulo notation, the result (10.9) of \cite{Friedlander}. However, repeating similar reasoning in the case of (p-p) cyclidic coordinates we arrive at the result different from (11.6) of \cite{Friedlander}. The details of such calculations will be published in the extended version of this paper.

\end{document}